\documentclass[
reprint,
superscriptaddress,
aps,
prl,
]{revtex4-2}

\usepackage{graphicx}
\usepackage{dcolumn}
\usepackage{bm}
\usepackage[utf8]{inputenc}
\usepackage{amsmath, amsthm, amssymb, amsfonts}
\usepackage{accents}
\usepackage[usenames,dvipsnames]{xcolor}
\usepackage{tikz}
\usepackage{array}
\usepackage{dsfont}
\usepackage{mathtools}
\usepackage{soul,xcolor}
\usepackage[colorlinks=true,linkcolor=Blue,citecolor=Blue,urlcolor=Blue]{hyperref}
\usepackage[caption=false]{subfig}
\usepackage[capitalise]{cleveref}

\newcommand{\YZ}{\color{black}}

\setstcolor{blue}

\begin{document}

\newcommand{\mytitle}{Mechanism for Strong Chimeras} 

\title{\mytitle}

\author{Yuanzhao Zhang}
\affiliation{Department of Physics and Astronomy, Northwestern University, Evanston, Illinois 60208, USA}
\author{Adilson E. Motter}
\affiliation{Department of Physics and Astronomy, Northwestern University, Evanston, Illinois 60208, USA}
\affiliation{Northwestern Institute on Complex Systems, Northwestern University, Evanston, Illinois 60208, USA}


\begin{abstract}
Chimera states have attracted significant attention as symmetry-broken states exhibiting the unexpected coexistence of coherence and incoherence. Despite the valuable insights gained from analyzing specific systems, an understanding of the general physical mechanism underlying the emergence of chimeras is still lacking. Here, we show that many stable chimeras arise because coherence in part of the system is sustained by incoherence in the rest of the system. This mechanism may be regarded as a deterministic analog of noise-induced synchronization and is shown to underlie the emergence of {\it strong} chimeras. These are chimera states whose coherent domain is formed by identically synchronized oscillators. Recognizing this mechanism offers a new meaning to the interpretation that chimeras are a natural link between coherence and incoherence.
\vspace{3mm}

\noindent DOI: \href{https://doi.org/10.1103/PhysRevLett.126.094101}{10.1103/PhysRevLett.126.094101} 
\end{abstract}

\maketitle

Chimera states are a remarkable phenomenon in which coherence and incoherence coexist in a system of identically-coupled identical oscillators \cite{panaggio2015chimera,omel2018mathematics}.
Initially regarded as a state that requires specific nonlocal coupling structure \cite{kuramoto2002coexistence,kuramoto2003nonlinear,abrams2004chimera} and/or specially prepared initial conditions \cite{abrams2008solvable,martens2010bistable}, chimeras have since been shown to be a general phenomenon that can occur robustly as a system (upon parameter changes) transitions from coherence to incoherence \cite{omel2008chimera,bordyugov2010self,hagerstrom2012experimental,sethia2013amplitude,schmidt2014coexistence,martens2016basins,bick2017robust}.
{\YZ Moreover, chimeras have been observed in a diverse set of physical systems, including those of optoelectronic \cite{hagerstrom2012experimental,larger2013virtual,hart2016experimental,zhang2020critical}, electrochemical \cite{tinsley2012chimera,schmidt2014coexistence,bick2017robust,totz2018spiral}, and mechanical \cite{martens2013chimera} nature.
There is even evidence pointing to chimera-like states in quantum systems \cite{bastidas2015quantum} and in the brain \cite{bansal2019cognitive}.}
Despite numerous efforts to elucidate the underlying principles \cite{martens2013chimera,sethia2014chimera,yeldesbay2014chimeralike,semenova2015does,schmidt2015clustering,semenova2016coherence,nicolaou2017chimera,kotwal2017connecting}, currently no system-independent mechanistic explanation exists that can provide broad physical insight into the emergence of chimeras.

Our goal is to bridge this gap by proposing a general mechanism for chimeras that is not tied to specific node dynamics, network structure, or coupling scheme.
We consider the important class of {\it permanently stable} chimera states whose coherent part is {\it identically} synchronized, which have been observed for both periodic oscillators \cite{yeldesbay2014chimeralike,schmidt2014coexistence,panaggio2016chimera} and chaotic oscillators \cite{hart2016experimental,cho2017stable,zhang2020critical}.
We also focus on parameter regions where global coherence is {\it unstable}, so the chimeras may be observed without the need of specially prepared initial conditions \cite{omel2008chimera,sieber2014controlling,yeldesbay2014chimeralike,schmidt2014coexistence,schmidt2015clustering,kotwal2017connecting,cho2017stable,totz2018spiral,zhang2020critical}.
Here, chimera states that 
(i) are permanently stable, 
(ii) exhibit identically synchronized coherent domain, and 
(iii) do not co-occur with stable global synchronization 
are referred to as {\it strong} chimeras. 
{\YZ Such states represent a diverse range of chimeras and have been observed in myriad systems \cite{yeldesbay2014chimeralike,schmidt2014coexistence,schmidt2015clustering,cho2017stable,zhang2020critical}}.

\begin{figure}[t]
\centering
\subfloat[]{
\includegraphics[width=\columnwidth]{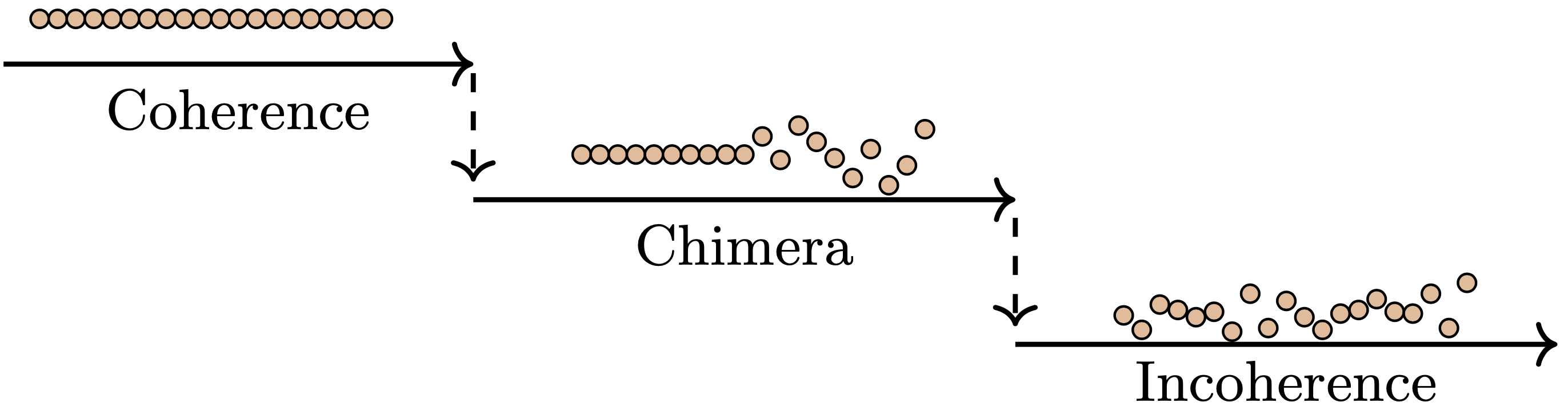}
}
\vspace{-4mm}
\caption{Example scenario considered in this study. 
As a bifurcation parameter is varied and the system transitions from coherence to incoherence, an intermediate chimera region emerges \cite{github}. 
}
\label{fig:1}
\end{figure}

In this Letter, we characterize strong chimeras that emerge between a globally synchronized state and a globally incoherent state as a bifurcation parameter is varied (\cref{fig:1}).
In such chimera states, the coexistence of a synchronous and an incoherent cluster challenges the intuition that inputs from the incoherent cluster would inevitably desynchronize the other cluster. 
Yet, our analysis shows that incoherence in part of the system in fact stabilizes the otherwise unstable coherence in the rest of the system, thus preventing a direct transition from global coherence to global incoherence when the former becomes unstable.
This incoherence-stabilized coherence effect is in many ways analogous to synchronization induced by common noise \cite{zhou2002noise,goldobin2005synchronization,nagai2010noise,pimenova2016interplay} and serves as a general mechanism giving rise to strong chimeras.

We first note that strong chimeras can be observed for both periodic and chaotic node dynamics, and in networks with either diffusive or non-diffusive coupling. 
In all cases, the necessary and sufficient condition for a cluster to admit an identical synchronization solution is that each of its oscillators receives the same input from the network. 
{\YZ This condition translates generically to the coherent cluster belonging to an equitable partition \cite{belykh2011mesoscale} and not being intertwined \cite{pecora2014cluster} with the rest of the network, although intertwined clusters are still allowed elsewhere in the network.}
(For diffusive coupling, the partition can be further relaxed to be {\it externally} equitable \cite{schaub2016graph}.)
The conditions above extend immediately to strong chimeras consisting of multiple coherent and incoherent clusters.

The impact of the rest of the network on the coherent cluster of a strong chimera can be analyzed by considering a network of $N$ coupled oscillators described by
\begin{equation}
x_i^{t+1} = \beta f(x_i^t) + K \sum_{j=1}^{N} M_{ij} h(x_j^t), \quad i=1,\dots,N,
\label{eq:full_net_dyn}
\end{equation}
where $x_i^t$ is the state of the $i$th oscillator at time $t$, function $f$ governs the dynamics of the uncoupled oscillators, $\beta$ denotes the self-feedback strength of the oscillators, $\bm{M}=(M_{ij})$ is the coupling matrix representing the network structure, $h$ is the coupling function, and $K$ is the overall coupling strength. 
We assume the oscillators to be time discrete and one dimensional for simplicity, but the analysis extends straightforwardly to continuous-time and higher-dimensional systems.
The matrix $\bm{M}$ can be rather general, including both diffusive and non-diffusive coupling schemes. 
Now, suppose that $C$ is the coherent cluster and that it consists of $n$ nodes numbered from $1$ to $n$. For oscillators in this cluster, the dynamical equation takes the form
\begin{equation}
x_i^{t+1}  = \beta f(x_i^t) + K \sum_{j=1}^{n} M_{ij} h(x_j^t) + I(t),  \quad i=1,\dots,n,
\label{eq:cluster_dyn}
\end{equation}
where $I(t)=K \sum_{j=n+1}^{N} M_{ij} h(x_j^t)$ is the input from the rest of the network, which does not depend on $i$ since the cluster must belong to a partition that is at least externally equitable.
Thus, the function $I(t)$ is common across all nodes in $C$, and the identical synchronization state $s^t$ in this cluster is given by
\begin{equation}
s^{t+1} = \beta f\big(s^t\big) + K\mu\,h\big(s^t\big) + I(t),
\label{eq:sync_trj}
\end{equation}
where {\YZ the row sum} $\mu = \sum_{j=1}^{n} M_{ij}$ is a constant not depending on $i$ for any $1\leq i \leq n$. 
The stability of this state is determined by the largest transverse Lyapunov exponent (LTLE) $\Lambda$ specified by the variational equations
\begin{equation}
\eta_i^{t+1} = \Big[\beta f'(s^t) + K\widehat{\lambda}_i h'(s^t) \Big] \eta_i^t, \quad i=2,\dots,n,
\label{eq:var_eq}
\end{equation}
where $\widehat{\lambda}_i$ is the $i$th eigenvalue of the $n\times n$ sub-coupling matrix $\widehat{\bm{M}} = (M_{ij})_{1\leq i,j \leq n}$ and $\eta_i$ is the corresponding perturbation mode.
The mode associated with $\widehat{\lambda}_1=\mu$ is excluded, as it corresponds to perturbations parallel to the synchronization manifold. 
Equation~(\ref{eq:var_eq}) implicitly assumes that $\widehat{\bm{M}}$ is diagonalizable, but this assumption can be lifted using the Jordan canonical form of this matrix \cite{nishikawa2006maximum,hart2019topological}.

We explicitly examine {\YZ \cref{eq:full_net_dyn} for} the two most widely studied coupling schemes, namely, diffusive coupling defined by the Laplacian matrix $\bm{L}$ and non-diffusive coupling defined by the adjacency matrix $\bm{A}$. 
For Laplacian coupling, $\bm{M} = -\bm{L}$ and, thus, $\mu$ is the negative of the indegree of nodes in $C$ due to connections from the rest of the network.
The eigenvalues $\widehat{\lambda}_i$ are given by $\widehat{\lambda}_i = - \lambda_i + \mu$, where $\lambda_i$ are the eigenvalues of the Laplacian
matrix of $C$ in isolation (i.e., consisting of intracluster connections only). 
For adjacency-matrix coupling,  $\bm{M} = \bm{A}$, the factor $\mu$ is the indegree of nodes in $C$ when the cluster is considered in isolation, and the eigenvalues are $\widehat{\lambda}_i = \lambda_i$, where $\lambda_i$ are the eigenvalues of the adjacency matrix of $C$ in isolation. 

\begin{figure*}[bt]
\centering
\includegraphics[width=1.9\columnwidth]{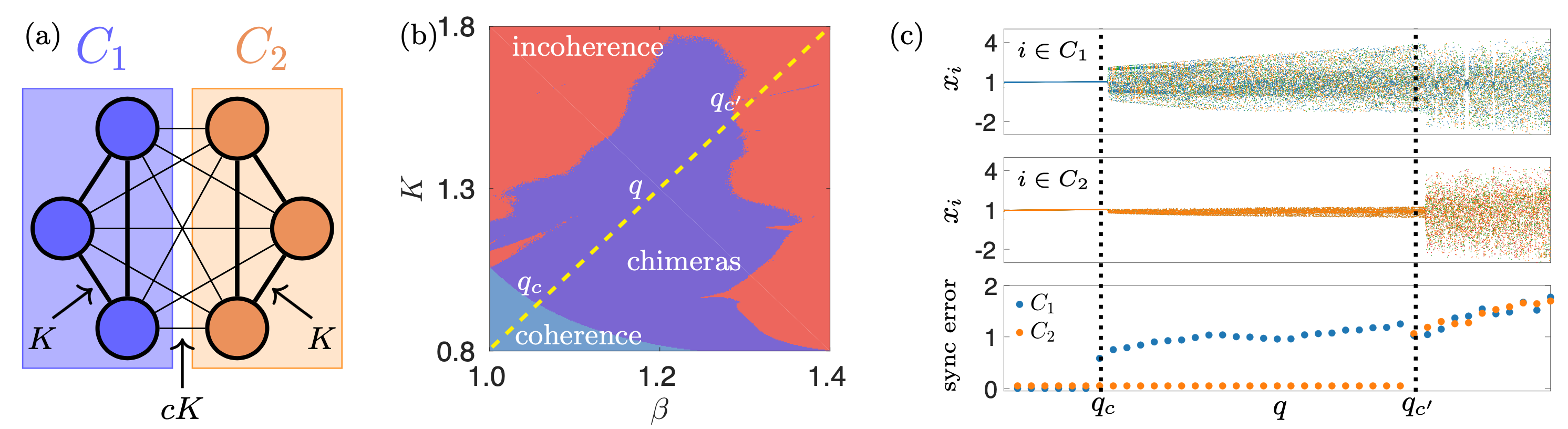} 
\vspace{-4mm}
\caption{Strong chimeras in a network of diffusively coupled  oscillators given by Eq.~(\ref{eq:1}).
(a) Network consisting of two identical clusters, $C_1$ and $C_2$. 
(b) Diagram in the  $\beta-K$ parameter space marking the  regions for 
which the system exhibits coherence (cyan), chimeras (purple), and incoherence (red) for $c=0.2$. 
(c) State transitions as parameter $q$ is varied quasistatically along the dashed line in (b) \cite{SM1}, showing the abrupt change from coherence to a chimera at $q_c$ and then from the chimera to incoherence at $q_{c'}$.
{\YZ The top and middle panels show the states $x_i^t$ in each cluster colored by the individual oscillators, and the bottom panel shows the corresponding time-averaged synchronization errors. The error is defined by the standard deviation among the oscillator states.}
}
\label{fig:2}
\end{figure*}

We first consider Laplacian coupling and, for concreteness, 
focus on networks composed of two {\it identical} clusters with all-to-all intercluster coupling. 
These networks are known to exhibit chimera states, which have been extensively studied in the literature \cite{abrams2008solvable,tinsley2012chimera,martens2013chimera,panaggio2016chimera}.
Assume that, as a bifurcation parameter $q$ is increased, the coherent state of a two-cluster network becomes unstable at a critical value $q_c$.
Because this point marks the end of coherence, at least one cluster must become incoherent  
when $q$ is further increased (see Supplemental Material \cite{SM} for details). 
Since the two clusters are identical, one might expect that both will become unstable at $q_c$ and that 
the system will thus 
transition directly from coherence (both clusters synchronized) into incoherence (both clusters incoherent).
That is, due to the symmetry between the two clusters, both clusters are expected to lose synchrony at the same time.
Nevertheless, chimera states often emerge right at the instability transition, breaking the symmetry between the clusters, with only one cluster becoming incoherent while the other  remains perfectly synchronized.
So, what prevents the system from evolving directly into global incoherence? 
The short answer is that, beyond $q_c$, incoherence in one cluster {\YZ can stabilize} synchronization in the other cluster, delaying the onset of global incoherence and instead giving rise to a chimera.

To further investigate this question, we focus on the node dynamics and coupling function given by
\begin{equation}
  f(x) = h(x) = \sin^2(x+\pi/4), 
  \label{eq:1}
\end{equation}
which model optoelectronic oscillators that have been realized in synchronization experiments \cite{hart2017experiments,zhang2020critical}.
While the intracluster coupling structure can be arbitrary in general, for clarity we focus on a network consisting of two clusters of $n=3$ nodes.
The clusters have internal coupling of strength $K$ and are connected to each other by all-to-all coupling of strength $cK$ [\cref{fig:2}(a)], where $c$ is included in matrix $\bm{M}$ in the representation of Eq.~(\ref{eq:full_net_dyn}).
There is nothing special about this choice of dynamics and cluster structure, and {\YZ we show in Supplemental Material \cite{SM} that our conclusions hold for other oscillators (with both discrete and continuous dynamics) and networks.
There, we also demonstrate the robustness of our results against oscillator heterogeneity.}

\Cref{fig:2}(b) shows the corresponding state diagram in the $\beta-K$ parameter space for $c=0.2$.
The classification of states in the diagram is based on the linear stability analysis of the coherent  and  chimera states as determined by the corresponding LTLE \cite{zhang2020critical}.
A generic bifurcation scenario is depicted in \cref{fig:2}(c): As a linear combination of the parameters $\beta$ and  $K$ is increased [dashed line in \cref{fig:2}(b)],
the system transitions from global coherence to a chimera state, and then from the chimera state to global incoherence.
In this example, the  chimera is defined by incoherence in cluster $C_1$ and coherence in cluster $C_2$. 
Starting from random initial conditions, it is equally likely for the clusters to exhibit swapped states, corresponding to a
chimera in $q_c<q<q_{c'}$ that has incoherence in $C_2$ and coherence in $C_1$. 

We can now establish a theoretical foundation for the mechanism underlying the onset of such chimeras by examining Eqs.~(\ref{eq:sync_trj}) and (\ref{eq:var_eq}). 
Crucially, the input from the rest of the network is irregular temporally but uniform spatially and does not affect the variational equations of $C$ directly, since $I(t)$ does not appear in Eq.~(\ref{eq:var_eq}).
Yet, it indirectly impacts synchronization stability by changing the synchronous state $s^t$ according to Eq.~(\ref{eq:sync_trj}).
It is entirely through the change it causes to $s^t$ that incoherence in the rest of the network stabilizes coherence in $C$, giving rise to a stable chimera.

\begin{figure}[b]  
\centering
\subfloat[]{
\includegraphics[width=0.99\columnwidth]{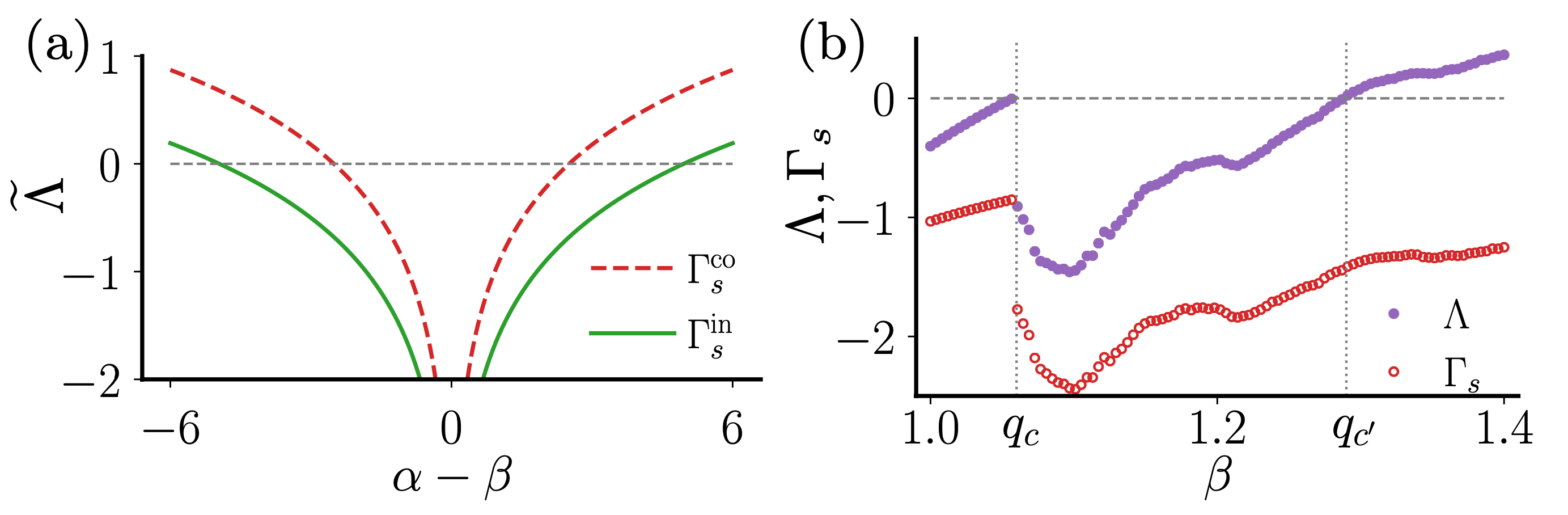} 
}
\vspace{-6mm} 
\caption{Impact of the incoherent cluster on the stability of the coherent one. 
(a) MSF for the coherent cluster before (red) and after (green) the other cluster transitions to incoherence, showing a widening of the stable region.
(b) LTLE of the coherent cluster (purple) as the parameter $q=q(\beta)$ in \cref{fig:2}(b) is varied, showing a discontinuous transition at $q_c$ and a continuous one at $q_{c'}$ due to the corresponding changes in $\Gamma_s$ (red). The system and other parameters are as in \cref{fig:2}.
}
\label{fig:3}
\end{figure}

To establish this rigorously, we note that the Lyapunov exponents of Eq.~(\ref{eq:var_eq}) can be written as $\Lambda^{(i)}=\ln |-K\widehat{\lambda}_i -\beta| +\Gamma_s$, where $\Gamma_s=\lim_{T\rightarrow \infty} {\frac{1}{T}} \ln \left| \Pi_{t=1}^{T} f'(s^t)\right |$ for $f'(x)= h'(x)$ as in the systems explicitly examined here.
Since $\Gamma_s$ is generally finite, the associated master stability function (MSF) \cite{pecora1998master} $\widetilde{\Lambda} (\alpha, \beta)=\ln |\alpha -\beta| +\Gamma_s$ defines a finite stability region, and synchronization in $C$ is stable if and only if 
\begin{equation}
|-K\widehat{\lambda}_i-\beta|<e^{-\Gamma_s}, \quad i=2,\dots,n.
\label{eq:stab}
\end{equation}
This equation explains what happens at the interface between global coherence and a chimera state.
As the bifurcation parameter $q$ is varied and the desynchronization transition is approached from the left, the condition in Eq.~(\ref{eq:stab}) is violated by at least one transverse mode and $\Lambda=\max_{i\ge 2}\{\Lambda^{(i)}\}$ vanishes at $q= q_c$.
But past this point, desynchronization in the other clusters alters $s^t$ dramatically, and $\Gamma_s$ changes (abruptly) from $\Gamma_s^{co}$ to $\Gamma_s^{in}$ according to its dependence on Eq.~(\ref{eq:sync_trj}).  
If $\Gamma_s^{in}<\Gamma_s^{co}$, the stability region defined by Eq.~(\ref{eq:stab}) expands (i.e., $\Lambda$ becomes negative again for $q>q_c$), and a chimera region then emerges due to stabilization caused by the incoherent input $I(t)$. 
In the example in \cref{fig:2}(b), in particular, $\Gamma_s^{co}=-0.92$ and $\Gamma_s^{in}=-1.6$, confirming our phenomenological observation that incoherence in $C_1$ stabilizes coherence in $C_2$.
\Cref{fig:3} shows the impact of this change on $\widetilde{\Lambda}$ [\cref{fig:3}(a)] as well as the codependence of $\Lambda$ and $\Gamma_s$ as the bifurcation parameter is varied [\cref{fig:3}(b)].

\begin{figure}[tb]  
\centering
\subfloat[]{
\includegraphics[width=\columnwidth]{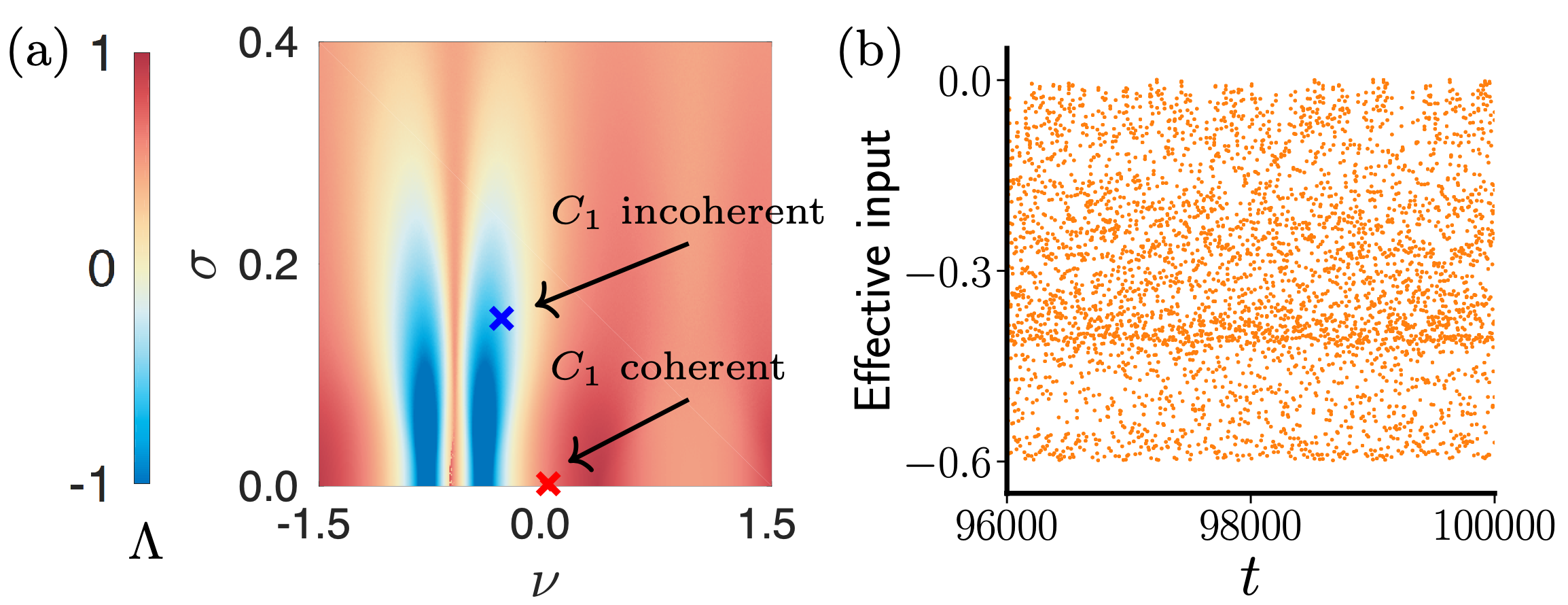}
}
\vspace{-6mm}
\caption{
Incoherence-stabilized coherence with the influence of the incoherent cluster modeled as a common noise driving term.
The system is the same as in \cref{fig:2} for $\beta=1.2$, $K=1.2$, and $C_2$ as the coherent cluster.
(a) LTLE of the coherent cluster for $c=0.2$ and the driving signal drawn from a Gaussian distribution of mean $\nu$ and standard deviation $\sigma$.
Also marked are the corresponding ($\nu$, $\sigma$) of the effective input for the original system [\cref{eq:sync_trj}] when the other cluster is incoherent (blue cross) or coherent (red cross).
{\YZ (b) Typical time series of the effective input corresponding to the blue cross.}
}
\label{fig:4}
\end{figure}

To further validate the hypothesis that the synchronization stability in the coherent cluster can be induced by the incoherent driving, we model the {\it effective} input $K\mu\,h\big(s^t\big) + I(t)$ in Eq.~(\ref{eq:sync_trj}) as a common noise driving term $\xi(t)$.
The synchronization trajectory $s^t$ in the coherent cluster is then 
\begin{equation}
	s^{t+1} = \beta f\big(s^t\big) + \xi(t),
	\label{eq:2}
\end{equation}
with the corresponding variational equations given by \cref{eq:var_eq} for $\widehat{\lambda}_i = -\lambda_i-cn$.
\Cref{fig:4}(a) shows the result of our stability analysis for $\beta=1.2$, $K=1.2$, and $c=0.2$ {\YZ (see Supplemental Material \cite{SM} for results on other values of $c$)}.
We see that synchronization in the coherent cluster is unstable ($\Lambda>0$) in the absence of external driving {\YZ $\xi(t)$}, but it can be stabilized ($\Lambda<0$) by a Gaussian {\YZ white} noise over a range of mean $\nu$ and standard deviation $\sigma$.
Moreover, the effective input from direct simulations of the incoherent cluster has $\nu=-0.30$ and $\sigma=0.15$, which is inside the stable region in \cref{fig:4}(a).
{\YZ \Cref{fig:4}(b) shows a typical time series of the effective input, which indeed closely resembles a noisy signal.
Our finding shares similarities with noise-induced synchronization, whose connection to chimera states was already speculated for specific systems, such as globally-coupled Kuramoto-Sakaguchi oscillators with delayed self-feedback \cite{yeldesbay2014chimeralike}.
However, it is worth noting that certain negative $\nu$ can stabilize coherence even when $\sigma=0$, which can potentially explain chimera death \cite{zakharova2014chimera}.}
On the other hand, the scenario in which both clusters are synchronized corresponds to $\nu=\sigma=0$ for diffusive coupling, which is equivalent to not having an external driving signal, and is unstable for the given parameters.
More generally, for diffusive coupling, coherence in one cluster can benefit from common driving only when the other cluster is not in the same state.

\begin{figure*}[tb] 
\centering
\subfloat[]{
\includegraphics[width=\textwidth]{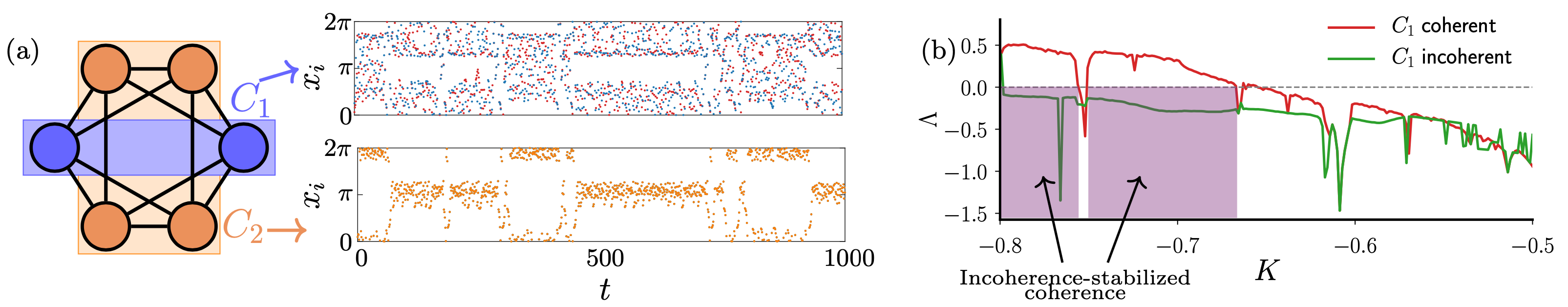} 
}
\vspace{-6mm} 
\caption{Strong chimeras in a network of non-diffusively coupled oscillators given by \cref{eq:41}.
(a) Network of identically-coupled oscillators organized into an incoherent ($C_1$) and a coherent ($C_2$) cluster.
Right: representative chimera trajectory for $K=-0.72$.
(b) LTLE of the coherent cluster as a function of $K$, where the other cluster is taken to be in a coherent state (red) or an incoherent one (green).
In the shaded region, coherence in $C_2$ is stabilized by incoherence in $C_1$.
}
\label{fig:5}
\end{figure*}

We now turn to adjacency-matrix coupling for
\begin{equation}
f(x) =  h(x) +(\pi/6)/\beta = [1-\cos(x)]/2 +(\pi/6)/\beta, 
  \label{eq:41}
\end{equation}
which is a closely related class of optoelectronic oscillators for which this type of coupling has been implemented experimentally \cite{pecora2014cluster}.
Here, the dynamical variables are constrained  to the interval $[0,2\pi)$ by taking $\text{mod} \, 2\pi$ at each iteration.
Our primary goal with this model is to illustrate a coupling scheme for which the intercluster coupling term does not vanish in the coherent state.
But we also want to show that our results do not depend on the coherent and incoherent clusters being equal. 
To facilitate comparison with the literature, we adopt a network and parameter setting for the system in Eq.~(\ref{eq:41}) first considered in Ref.~\cite{cho2017stable}.
The network consists of a ring of six nodes coupled to their first- and second-nearest neighbors [\cref{fig:5}(a)], and the parameters are set to $\beta = 2\pi/3 - 4K$.

In \cref{fig:5}(b), we perform a comparative analysis and plot the LTLE for coherence in cluster $C_2$ (comprising four oscillators) when cluster $C_1$ (comprising two oscillators) is assumed to be coherent and incoherent, respectively.
We see that incoherence in $C_1$ significantly delays the instability transition in $C_2$ from $K=-0.67$ to $K=-0.80$ and, as a consequence, gives rise to a much wider chimera region than previously expected \cite{cho2017stable}. 
A representative time series for the chimera state at $K=-0.72$, whose coherent cluster is stabilized by the incoherent one, is shown in \cref{fig:5}(a).

The analysis presented above reveals a physical mechanism underlying the emergence of strong chimeras.  
The self-consistency of such states was  partially elucidated by the previous demonstration that desynchronization in one cluster does not necessarily lead to the concurrent desynchronization in another cluster  \cite{pecora2014cluster,cho2017stable}. 
Here, we have been able to go one step further and demonstrate that incoherence in one cluster can, in fact, stabilize coherence in the other cluster. 
This incoherence-stabilized coherence adds a new dimension to the proposition that chimera states are the natural link between coherent and incoherent states \cite{omel2008chimera,omelchenko2011loss,omelchenko2012transition}.

{\YZ Our results have potential implications for seizures that arise from the excessive synchronization of large neuronal populations \cite{jiruska2013synchronization}, which have been linked to chimera states \cite{rothkegel2014irregular,andrzejak2016all,chouzouris2018chimera,lainscsek2019cortical}. 
In particular, the discovery that the coherent domain is often stabilized by its interaction with the incoherent domain can help explain why corpus callosotomy \cite{asadi2008corpus}, surgery that separates the brain into two disconnected hemispheres, is an effective treatment for epilepsy and seizures.
Incoherence-stabilized coherence also provides insights on the counter-intuitive phenomena that desynchronization is often observed preceding seizures \cite{netoff2002decreased,mormann2003epileptic,andrzejak2016all} and that high levels of synchrony can facilitate seizure termination \cite{schindler2007assessing,lainscsek2019cortical}.
}

As a promising direction for future research, we note that chimera states not meeting the conditions for strong chimeras have been studied in the literature, including those with coherent domains that are not identically synchronized \cite{abrams2004chimera,hagerstrom2012experimental,bick2017robust,bogomolov2017mechanisms} and those that co-occur with stable global coherence \cite{abrams2008solvable,martens2010bistable,sethia2014chimera,kotwal2017connecting}.
It remains to be shown how the mechanism uncovered for strong chimeras may provide insight into those states.
{\YZ The key challenge is that oscillators in the coherent domain are typically not driven by a common input signal when they are not identically synchronized, although they often remain approximately synchronized and receive similar input \cite{abrams2004chimera,omelchenko2011loss}.}
Given the prevalence of related phenomena such as noise-induced synchronization~\cite{zhou2002noise2,teramae2004robustness,nakao2007noise,ullner2009noise}, 
we believe the cooperative relation between incoherence and coherence revealed by our analysis can be a general mechanism giving rise to a wide range of chimera states.

\begin{acknowledgments}
The authors thank Z.\ G.\ Nicolaou, J.\ D.\ Hart, and R.\ Roy for insightful discussions.
This work was supported by ARO Grant No.\ W911NF-19-1-0383.
\end{acknowledgments}

\bibliography{net_dyn}

\begin{thebibliography}{63}%
\makeatletter
\providecommand \@ifxundefined [1]{%
 \@ifx{#1\undefined}
}%
\providecommand \@ifnum [1]{%
 \ifnum #1\expandafter \@firstoftwo
 \else \expandafter \@secondoftwo
 \fi
}%
\providecommand \@ifx [1]{%
 \ifx #1\expandafter \@firstoftwo
 \else \expandafter \@secondoftwo
 \fi
}%
\providecommand \natexlab [1]{#1}%
\providecommand \enquote  [1]{``#1''}%
\providecommand \bibnamefont  [1]{#1}%
\providecommand \bibfnamefont [1]{#1}%
\providecommand \citenamefont [1]{#1}%
\providecommand \href@noop [0]{\@secondoftwo}%
\providecommand \href [0]{\begingroup \@sanitize@url \@href}%
\providecommand \@href[1]{\@@startlink{#1}\@@href}%
\providecommand \@@href[1]{\endgroup#1\@@endlink}%
\providecommand \@sanitize@url [0]{\catcode `\\12\catcode `\$12\catcode
  `\&12\catcode `\#12\catcode `\^12\catcode `\_12\catcode `\%12\relax}%
\providecommand \@@startlink[1]{}%
\providecommand \@@endlink[0]{}%
\providecommand \url  [0]{\begingroup\@sanitize@url \@url }%
\providecommand \@url [1]{\endgroup\@href {#1}{\urlprefix }}%
\providecommand \urlprefix  [0]{URL }%
\providecommand \Eprint [0]{\href }%
\providecommand \doibase [0]{https://doi.org/}%
\providecommand \selectlanguage [0]{\@gobble}%
\providecommand \bibinfo  [0]{\@secondoftwo}%
\providecommand \bibfield  [0]{\@secondoftwo}%
\providecommand \translation [1]{[#1]}%
\providecommand \BibitemOpen [0]{}%
\providecommand \bibitemStop [0]{}%
\providecommand \bibitemNoStop [0]{.\EOS\space}%
\providecommand \EOS [0]{\spacefactor3000\relax}%
\providecommand \BibitemShut  [1]{\csname bibitem#1\endcsname}%
\let\auto@bib@innerbib\@empty
\bibitem [{\citenamefont {Panaggio}\ and\ \citenamefont
  {Abrams}(2015)}]{panaggio2015chimera}%
  \BibitemOpen
  \bibfield  {author} {\bibinfo {author} {\bibfnamefont {M.~J.}\ \bibnamefont
  {Panaggio}}\ and\ \bibinfo {author} {\bibfnamefont {D.~M.}\ \bibnamefont
  {Abrams}},\ }\bibfield  {title} {\bibinfo {title} {Chimera states:
  Coexistence of coherence and incoherence in networks of coupled
  oscillators},\ }\href@noop {} {\bibfield  {journal} {\bibinfo  {journal}
  {Nonlinearity}\ }\textbf {\bibinfo {volume} {28}},\ \bibinfo {pages} {R67}
  (\bibinfo {year} {2015})}\BibitemShut {NoStop}%
\bibitem [{\citenamefont {Omel'chenko}(2018)}]{omel2018mathematics}%
  \BibitemOpen
  \bibfield  {author} {\bibinfo {author} {\bibfnamefont {O.~E.}\ \bibnamefont
  {Omel'chenko}},\ }\bibfield  {title} {\bibinfo {title} {The mathematics
  behind chimera states},\ }\href@noop {} {\bibfield  {journal} {\bibinfo
  {journal} {Nonlinearity}\ }\textbf {\bibinfo {volume} {31}},\ \bibinfo
  {pages} {R121} (\bibinfo {year} {2018})}\BibitemShut {NoStop}%
\bibitem [{\citenamefont {Kuramoto}\ and\ \citenamefont
  {Battogtokh}(2002)}]{kuramoto2002coexistence}%
  \BibitemOpen
  \bibfield  {author} {\bibinfo {author} {\bibfnamefont {Y.}~\bibnamefont
  {Kuramoto}}\ and\ \bibinfo {author} {\bibfnamefont {D.}~\bibnamefont
  {Battogtokh}},\ }\bibfield  {title} {\bibinfo {title} {Coexistence of
  coherence and incoherence in nonlocally coupled phase oscillators},\
  }\href@noop {} {\bibfield  {journal} {\bibinfo  {journal} {Nonlinear Phenom.
  Complex Syst.}\ }\textbf {\bibinfo {volume} {5}},\ \bibinfo {pages} {380}
  (\bibinfo {year} {2002})}\BibitemShut {NoStop}%
\bibitem [{\citenamefont {Kuramoto}(2002)}]{kuramoto2003nonlinear}%
  \BibitemOpen
  \bibfield  {author} {\bibinfo {author} {\bibfnamefont {Y.}~\bibnamefont
  {Kuramoto}},\ }\href@noop {} {\emph {\bibinfo {title} {Nonlinear Dynamics and
  Chaos: Where Do We Go from Here?}}}\ (\bibinfo  {publisher} {CRC Press},\
  \bibinfo {year} {2002})\ pp.\ \bibinfo {pages} {209--227}\BibitemShut
  {NoStop}%
\bibitem [{\citenamefont {Abrams}\ and\ \citenamefont
  {Strogatz}(2004)}]{abrams2004chimera}%
  \BibitemOpen
  \bibfield  {author} {\bibinfo {author} {\bibfnamefont {D.~M.}\ \bibnamefont
  {Abrams}}\ and\ \bibinfo {author} {\bibfnamefont {S.~H.}\ \bibnamefont
  {Strogatz}},\ }\bibfield  {title} {\bibinfo {title} {Chimera states for
  coupled oscillators},\ }\href@noop {} {\bibfield  {journal} {\bibinfo
  {journal} {Phys. Rev. Lett.}\ }\textbf {\bibinfo {volume} {93}},\ \bibinfo
  {pages} {174102} (\bibinfo {year} {2004})}\BibitemShut {NoStop}%
\bibitem [{\citenamefont {Abrams}\ \emph {et~al.}(2008)\citenamefont {Abrams},
  \citenamefont {Mirollo}, \citenamefont {Strogatz},\ and\ \citenamefont
  {Wiley}}]{abrams2008solvable}%
  \BibitemOpen
  \bibfield  {author} {\bibinfo {author} {\bibfnamefont {D.~M.}\ \bibnamefont
  {Abrams}}, \bibinfo {author} {\bibfnamefont {R.}~\bibnamefont {Mirollo}},
  \bibinfo {author} {\bibfnamefont {S.~H.}\ \bibnamefont {Strogatz}},\ and\
  \bibinfo {author} {\bibfnamefont {D.~A.}\ \bibnamefont {Wiley}},\ }\bibfield
  {title} {\bibinfo {title} {Solvable model for chimera states of coupled
  oscillators},\ }\href@noop {} {\bibfield  {journal} {\bibinfo  {journal}
  {Phys. Rev. Lett.}\ }\textbf {\bibinfo {volume} {101}},\ \bibinfo {pages}
  {084103} (\bibinfo {year} {2008})}\BibitemShut {NoStop}%
\bibitem [{\citenamefont {Martens}(2010)}]{martens2010bistable}%
  \BibitemOpen
  \bibfield  {author} {\bibinfo {author} {\bibfnamefont {E.~A.}\ \bibnamefont
  {Martens}},\ }\bibfield  {title} {\bibinfo {title} {Bistable chimera
  attractors on a triangular network of oscillator populations},\ }\href@noop
  {} {\bibfield  {journal} {\bibinfo  {journal} {Phys. Rev. E}\ }\textbf
  {\bibinfo {volume} {82}},\ \bibinfo {pages} {016216} (\bibinfo {year}
  {2010})}\BibitemShut {NoStop}%
\bibitem [{\citenamefont {Omel'chenko}\ \emph {et~al.}(2008)\citenamefont
  {Omel'chenko}, \citenamefont {Maistrenko},\ and\ \citenamefont
  {Tass}}]{omel2008chimera}%
  \BibitemOpen
  \bibfield  {author} {\bibinfo {author} {\bibfnamefont {O.~E.}\ \bibnamefont
  {Omel'chenko}}, \bibinfo {author} {\bibfnamefont {Y.~L.}\ \bibnamefont
  {Maistrenko}},\ and\ \bibinfo {author} {\bibfnamefont {P.~A.}\ \bibnamefont
  {Tass}},\ }\bibfield  {title} {\bibinfo {title} {Chimera states: The natural
  link between coherence and incoherence},\ }\href@noop {} {\bibfield
  {journal} {\bibinfo  {journal} {Phys. Rev. Lett.}\ }\textbf {\bibinfo
  {volume} {100}},\ \bibinfo {pages} {044105} (\bibinfo {year}
  {2008})}\BibitemShut {NoStop}%
\bibitem [{\citenamefont {Bordyugov}\ \emph {et~al.}(2010)\citenamefont
  {Bordyugov}, \citenamefont {Pikovsky},\ and\ \citenamefont
  {Rosenblum}}]{bordyugov2010self}%
  \BibitemOpen
  \bibfield  {author} {\bibinfo {author} {\bibfnamefont {G.}~\bibnamefont
  {Bordyugov}}, \bibinfo {author} {\bibfnamefont {A.}~\bibnamefont
  {Pikovsky}},\ and\ \bibinfo {author} {\bibfnamefont {M.}~\bibnamefont
  {Rosenblum}},\ }\bibfield  {title} {\bibinfo {title} {Self-emerging and
  turbulent chimeras in oscillator chains},\ }\href@noop {} {\bibfield
  {journal} {\bibinfo  {journal} {Phys. Rev. E}\ }\textbf {\bibinfo {volume}
  {82}},\ \bibinfo {pages} {035205} (\bibinfo {year} {2010})}\BibitemShut
  {NoStop}%
\bibitem [{\citenamefont {Hagerstrom}\ \emph {et~al.}(2012)\citenamefont
  {Hagerstrom}, \citenamefont {Murphy}, \citenamefont {Roy}, \citenamefont
  {H{\"o}vel}, \citenamefont {Omelchenko},\ and\ \citenamefont
  {Sch{\"o}ll}}]{hagerstrom2012experimental}%
  \BibitemOpen
  \bibfield  {author} {\bibinfo {author} {\bibfnamefont {A.~M.}\ \bibnamefont
  {Hagerstrom}}, \bibinfo {author} {\bibfnamefont {T.~E.}\ \bibnamefont
  {Murphy}}, \bibinfo {author} {\bibfnamefont {R.}~\bibnamefont {Roy}},
  \bibinfo {author} {\bibfnamefont {P.}~\bibnamefont {H{\"o}vel}}, \bibinfo
  {author} {\bibfnamefont {I.}~\bibnamefont {Omelchenko}},\ and\ \bibinfo
  {author} {\bibfnamefont {E.}~\bibnamefont {Sch{\"o}ll}},\ }\bibfield  {title}
  {\bibinfo {title} {Experimental observation of chimeras in coupled-map
  lattices},\ }\href@noop {} {\bibfield  {journal} {\bibinfo  {journal} {Nat.
  Phys.}\ }\textbf {\bibinfo {volume} {8}},\ \bibinfo {pages} {658} (\bibinfo
  {year} {2012})}\BibitemShut {NoStop}%
\bibitem [{\citenamefont {Sethia}\ \emph {et~al.}(2013)\citenamefont {Sethia},
  \citenamefont {Sen},\ and\ \citenamefont {Johnston}}]{sethia2013amplitude}%
  \BibitemOpen
  \bibfield  {author} {\bibinfo {author} {\bibfnamefont {G.~C.}\ \bibnamefont
  {Sethia}}, \bibinfo {author} {\bibfnamefont {A.}~\bibnamefont {Sen}},\ and\
  \bibinfo {author} {\bibfnamefont {G.~L.}\ \bibnamefont {Johnston}},\
  }\bibfield  {title} {\bibinfo {title} {Amplitude-mediated chimera states},\
  }\href@noop {} {\bibfield  {journal} {\bibinfo  {journal} {Phys. Rev. E}\
  }\textbf {\bibinfo {volume} {88}},\ \bibinfo {pages} {042917} (\bibinfo
  {year} {2013})}\BibitemShut {NoStop}%
\bibitem [{\citenamefont {Schmidt}\ \emph {et~al.}(2014)\citenamefont
  {Schmidt}, \citenamefont {Sch{\"o}nleber}, \citenamefont {Krischer},\ and\
  \citenamefont {Garc{\'\i}a-Morales}}]{schmidt2014coexistence}%
  \BibitemOpen
  \bibfield  {author} {\bibinfo {author} {\bibfnamefont {L.}~\bibnamefont
  {Schmidt}}, \bibinfo {author} {\bibfnamefont {K.}~\bibnamefont
  {Sch{\"o}nleber}}, \bibinfo {author} {\bibfnamefont {K.}~\bibnamefont
  {Krischer}},\ and\ \bibinfo {author} {\bibfnamefont {V.}~\bibnamefont
  {Garc{\'\i}a-Morales}},\ }\bibfield  {title} {\bibinfo {title} {Coexistence
  of synchrony and incoherence in oscillatory media under nonlinear global
  coupling},\ }\href@noop {} {\bibfield  {journal} {\bibinfo  {journal}
  {Chaos}\ }\textbf {\bibinfo {volume} {24}},\ \bibinfo {pages} {013102}
  (\bibinfo {year} {2014})}\BibitemShut {NoStop}%
\bibitem [{\citenamefont {Martens}\ \emph {et~al.}(2016)\citenamefont
  {Martens}, \citenamefont {Panaggio},\ and\ \citenamefont
  {Abrams}}]{martens2016basins}%
  \BibitemOpen
  \bibfield  {author} {\bibinfo {author} {\bibfnamefont {E.~A.}\ \bibnamefont
  {Martens}}, \bibinfo {author} {\bibfnamefont {M.~J.}\ \bibnamefont
  {Panaggio}},\ and\ \bibinfo {author} {\bibfnamefont {D.~M.}\ \bibnamefont
  {Abrams}},\ }\bibfield  {title} {\bibinfo {title} {Basins of attraction for
  chimera states},\ }\href@noop {} {\bibfield  {journal} {\bibinfo  {journal}
  {New J. Phys.}\ }\textbf {\bibinfo {volume} {18}},\ \bibinfo {pages} {022002}
  (\bibinfo {year} {2016})}\BibitemShut {NoStop}%
\bibitem [{\citenamefont {Bick}\ \emph {et~al.}(2017)\citenamefont {Bick},
  \citenamefont {Sebek},\ and\ \citenamefont {Kiss}}]{bick2017robust}%
  \BibitemOpen
  \bibfield  {author} {\bibinfo {author} {\bibfnamefont {C.}~\bibnamefont
  {Bick}}, \bibinfo {author} {\bibfnamefont {M.}~\bibnamefont {Sebek}},\ and\
  \bibinfo {author} {\bibfnamefont {I.~Z.}\ \bibnamefont {Kiss}},\ }\bibfield
  {title} {\bibinfo {title} {Robust weak chimeras in oscillator networks with
  delayed linear and quadratic interactions},\ }\href@noop {} {\bibfield
  {journal} {\bibinfo  {journal} {Phys. Rev. Lett.}\ }\textbf {\bibinfo
  {volume} {119}},\ \bibinfo {pages} {168301} (\bibinfo {year}
  {2017})}\BibitemShut {NoStop}%
\bibitem [{\citenamefont {Larger}\ \emph {et~al.}(2013)\citenamefont {Larger},
  \citenamefont {Penkovsky},\ and\ \citenamefont
  {Maistrenko}}]{larger2013virtual}%
  \BibitemOpen
  \bibfield  {author} {\bibinfo {author} {\bibfnamefont {L.}~\bibnamefont
  {Larger}}, \bibinfo {author} {\bibfnamefont {B.}~\bibnamefont {Penkovsky}},\
  and\ \bibinfo {author} {\bibfnamefont {Y.}~\bibnamefont {Maistrenko}},\
  }\bibfield  {title} {\bibinfo {title} {Virtual chimera states for
  delayed-feedback systems},\ }\href@noop {} {\bibfield  {journal} {\bibinfo
  {journal} {Phys. Rev. Lett.}\ }\textbf {\bibinfo {volume} {111}},\ \bibinfo
  {pages} {054103} (\bibinfo {year} {2013})}\BibitemShut {NoStop}%
\bibitem [{\citenamefont {Hart}\ \emph {et~al.}(2016)\citenamefont {Hart},
  \citenamefont {Bansal}, \citenamefont {Murphy},\ and\ \citenamefont
  {Roy}}]{hart2016experimental}%
  \BibitemOpen
  \bibfield  {author} {\bibinfo {author} {\bibfnamefont {J.~D.}\ \bibnamefont
  {Hart}}, \bibinfo {author} {\bibfnamefont {K.}~\bibnamefont {Bansal}},
  \bibinfo {author} {\bibfnamefont {T.~E.}\ \bibnamefont {Murphy}},\ and\
  \bibinfo {author} {\bibfnamefont {R.}~\bibnamefont {Roy}},\ }\bibfield
  {title} {\bibinfo {title} {Experimental observation of chimera and cluster
  states in a minimal globally coupled network},\ }\href@noop {} {\bibfield
  {journal} {\bibinfo  {journal} {Chaos}\ }\textbf {\bibinfo {volume} {26}},\
  \bibinfo {pages} {094801} (\bibinfo {year} {2016})}\BibitemShut {NoStop}%
\bibitem [{\citenamefont {Zhang}\ \emph {et~al.}(2020)\citenamefont {Zhang},
  \citenamefont {Nicolaou}, \citenamefont {Hart}, \citenamefont {Roy},\ and\
  \citenamefont {Motter}}]{zhang2020critical}%
  \BibitemOpen
  \bibfield  {author} {\bibinfo {author} {\bibfnamefont {Y.}~\bibnamefont
  {Zhang}}, \bibinfo {author} {\bibfnamefont {Z.~G.}\ \bibnamefont {Nicolaou}},
  \bibinfo {author} {\bibfnamefont {J.~D.}\ \bibnamefont {Hart}}, \bibinfo
  {author} {\bibfnamefont {R.}~\bibnamefont {Roy}},\ and\ \bibinfo {author}
  {\bibfnamefont {A.~E.}\ \bibnamefont {Motter}},\ }\bibfield  {title}
  {\bibinfo {title} {Critical switching in globally attractive chimeras},\
  }\href@noop {} {\bibfield  {journal} {\bibinfo  {journal} {Phys. Rev. X}\
  }\textbf {\bibinfo {volume} {10}},\ \bibinfo {pages} {011044} (\bibinfo
  {year} {2020})}\BibitemShut {NoStop}%
\bibitem [{\citenamefont {Tinsley}\ \emph {et~al.}(2012)\citenamefont
  {Tinsley}, \citenamefont {Nkomo},\ and\ \citenamefont
  {Showalter}}]{tinsley2012chimera}%
  \BibitemOpen
  \bibfield  {author} {\bibinfo {author} {\bibfnamefont {M.~R.}\ \bibnamefont
  {Tinsley}}, \bibinfo {author} {\bibfnamefont {S.}~\bibnamefont {Nkomo}},\
  and\ \bibinfo {author} {\bibfnamefont {K.}~\bibnamefont {Showalter}},\
  }\bibfield  {title} {\bibinfo {title} {Chimera and phase-cluster states in
  populations of coupled chemical oscillators},\ }\href@noop {} {\bibfield
  {journal} {\bibinfo  {journal} {Nat. Phys.}\ }\textbf {\bibinfo {volume}
  {8}},\ \bibinfo {pages} {662} (\bibinfo {year} {2012})}\BibitemShut {NoStop}%
\bibitem [{\citenamefont {Totz}\ \emph {et~al.}(2018)\citenamefont {Totz},
  \citenamefont {Rode}, \citenamefont {Tinsley}, \citenamefont {Showalter},\
  and\ \citenamefont {Engel}}]{totz2018spiral}%
  \BibitemOpen
  \bibfield  {author} {\bibinfo {author} {\bibfnamefont {J.~F.}\ \bibnamefont
  {Totz}}, \bibinfo {author} {\bibfnamefont {J.}~\bibnamefont {Rode}}, \bibinfo
  {author} {\bibfnamefont {M.~R.}\ \bibnamefont {Tinsley}}, \bibinfo {author}
  {\bibfnamefont {K.}~\bibnamefont {Showalter}},\ and\ \bibinfo {author}
  {\bibfnamefont {H.}~\bibnamefont {Engel}},\ }\bibfield  {title} {\bibinfo
  {title} {Spiral wave chimera states in large populations of coupled chemical
  oscillators},\ }\href@noop {} {\bibfield  {journal} {\bibinfo  {journal}
  {Nat. Phys.}\ }\textbf {\bibinfo {volume} {14}},\ \bibinfo {pages} {282}
  (\bibinfo {year} {2018})}\BibitemShut {NoStop}%
\bibitem [{\citenamefont {Martens}\ \emph {et~al.}(2013)\citenamefont
  {Martens}, \citenamefont {Thutupalli}, \citenamefont {Fourri{\`e}re},\ and\
  \citenamefont {Hallatschek}}]{martens2013chimera}%
  \BibitemOpen
  \bibfield  {author} {\bibinfo {author} {\bibfnamefont {E.~A.}\ \bibnamefont
  {Martens}}, \bibinfo {author} {\bibfnamefont {S.}~\bibnamefont {Thutupalli}},
  \bibinfo {author} {\bibfnamefont {A.}~\bibnamefont {Fourri{\`e}re}},\ and\
  \bibinfo {author} {\bibfnamefont {O.}~\bibnamefont {Hallatschek}},\
  }\bibfield  {title} {\bibinfo {title} {Chimera states in mechanical
  oscillator networks},\ }\href@noop {} {\bibfield  {journal} {\bibinfo
  {journal} {Proc. Natl. Acad. Sci. U.S.A.}\ }\textbf {\bibinfo {volume}
  {110}},\ \bibinfo {pages} {10563} (\bibinfo {year} {2013})}\BibitemShut
  {NoStop}%
\bibitem [{\citenamefont {Bastidas}\ \emph {et~al.}(2015)\citenamefont
  {Bastidas}, \citenamefont {Omelchenko}, \citenamefont {Zakharova},
  \citenamefont {Sch{\"o}ll},\ and\ \citenamefont
  {Brandes}}]{bastidas2015quantum}%
  \BibitemOpen
  \bibfield  {author} {\bibinfo {author} {\bibfnamefont {V.~M.}\ \bibnamefont
  {Bastidas}}, \bibinfo {author} {\bibfnamefont {I.}~\bibnamefont
  {Omelchenko}}, \bibinfo {author} {\bibfnamefont {A.}~\bibnamefont
  {Zakharova}}, \bibinfo {author} {\bibfnamefont {E.}~\bibnamefont
  {Sch{\"o}ll}},\ and\ \bibinfo {author} {\bibfnamefont {T.}~\bibnamefont
  {Brandes}},\ }\bibfield  {title} {\bibinfo {title} {Quantum signatures of
  chimera states},\ }\href@noop {} {\bibfield  {journal} {\bibinfo  {journal}
  {Phys. Rev. E}\ }\textbf {\bibinfo {volume} {92}},\ \bibinfo {pages} {062924}
  (\bibinfo {year} {2015})}\BibitemShut {NoStop}%
\bibitem [{\citenamefont {Bansal}\ \emph {et~al.}(2019)\citenamefont {Bansal},
  \citenamefont {Garcia}, \citenamefont {Tompson}, \citenamefont {Verstynen},
  \citenamefont {Vettel},\ and\ \citenamefont {Muldoon}}]{bansal2019cognitive}%
  \BibitemOpen
  \bibfield  {author} {\bibinfo {author} {\bibfnamefont {K.}~\bibnamefont
  {Bansal}}, \bibinfo {author} {\bibfnamefont {J.~O.}\ \bibnamefont {Garcia}},
  \bibinfo {author} {\bibfnamefont {S.~H.}\ \bibnamefont {Tompson}}, \bibinfo
  {author} {\bibfnamefont {T.}~\bibnamefont {Verstynen}}, \bibinfo {author}
  {\bibfnamefont {J.~M.}\ \bibnamefont {Vettel}},\ and\ \bibinfo {author}
  {\bibfnamefont {S.~F.}\ \bibnamefont {Muldoon}},\ }\bibfield  {title}
  {\bibinfo {title} {Cognitive chimera states in human brain networks},\
  }\href@noop {} {\bibfield  {journal} {\bibinfo  {journal} {Sci. Adv.}\
  }\textbf {\bibinfo {volume} {5}},\ \bibinfo {pages} {eaau8535} (\bibinfo
  {year} {2019})}\BibitemShut {NoStop}%
\bibitem [{\citenamefont {Sethia}\ and\ \citenamefont
  {Sen}(2014)}]{sethia2014chimera}%
  \BibitemOpen
  \bibfield  {author} {\bibinfo {author} {\bibfnamefont {G.~C.}\ \bibnamefont
  {Sethia}}\ and\ \bibinfo {author} {\bibfnamefont {A.}~\bibnamefont {Sen}},\
  }\bibfield  {title} {\bibinfo {title} {Chimera states: The existence criteria
  revisited},\ }\href@noop {} {\bibfield  {journal} {\bibinfo  {journal} {Phys.
  Rev. Lett.}\ }\textbf {\bibinfo {volume} {112}},\ \bibinfo {pages} {144101}
  (\bibinfo {year} {2014})}\BibitemShut {NoStop}%
\bibitem [{\citenamefont {Yeldesbay}\ \emph {et~al.}(2014)\citenamefont
  {Yeldesbay}, \citenamefont {Pikovsky},\ and\ \citenamefont
  {Rosenblum}}]{yeldesbay2014chimeralike}%
  \BibitemOpen
  \bibfield  {author} {\bibinfo {author} {\bibfnamefont {A.}~\bibnamefont
  {Yeldesbay}}, \bibinfo {author} {\bibfnamefont {A.}~\bibnamefont
  {Pikovsky}},\ and\ \bibinfo {author} {\bibfnamefont {M.}~\bibnamefont
  {Rosenblum}},\ }\bibfield  {title} {\bibinfo {title} {Chimeralike states in
  an ensemble of globally coupled oscillators},\ }\href@noop {} {\bibfield
  {journal} {\bibinfo  {journal} {Phys. Rev. Lett.}\ }\textbf {\bibinfo
  {volume} {112}},\ \bibinfo {pages} {144103} (\bibinfo {year}
  {2014})}\BibitemShut {NoStop}%
\bibitem [{\citenamefont {Semenova}\ \emph {et~al.}(2015)\citenamefont
  {Semenova}, \citenamefont {Zakharova}, \citenamefont {Sch{\"o}ll},\ and\
  \citenamefont {Anishchenko}}]{semenova2015does}%
  \BibitemOpen
  \bibfield  {author} {\bibinfo {author} {\bibfnamefont {N.}~\bibnamefont
  {Semenova}}, \bibinfo {author} {\bibfnamefont {A.}~\bibnamefont {Zakharova}},
  \bibinfo {author} {\bibfnamefont {E.}~\bibnamefont {Sch{\"o}ll}},\ and\
  \bibinfo {author} {\bibfnamefont {V.}~\bibnamefont {Anishchenko}},\
  }\bibfield  {title} {\bibinfo {title} {Does hyperbolicity impede emergence of
  chimera states in networks of nonlocally coupled chaotic oscillators?},\
  }\href@noop {} {\bibfield  {journal} {\bibinfo  {journal} {Europhys. Lett.}\
  }\textbf {\bibinfo {volume} {112}},\ \bibinfo {pages} {40002} (\bibinfo
  {year} {2015})}\BibitemShut {NoStop}%
\bibitem [{\citenamefont {Schmidt}\ and\ \citenamefont
  {Krischer}(2015)}]{schmidt2015clustering}%
  \BibitemOpen
  \bibfield  {author} {\bibinfo {author} {\bibfnamefont {L.}~\bibnamefont
  {Schmidt}}\ and\ \bibinfo {author} {\bibfnamefont {K.}~\bibnamefont
  {Krischer}},\ }\bibfield  {title} {\bibinfo {title} {Clustering as a
  prerequisite for chimera states in globally coupled systems},\ }\href@noop {}
  {\bibfield  {journal} {\bibinfo  {journal} {Phys. Rev. Lett.}\ }\textbf
  {\bibinfo {volume} {114}},\ \bibinfo {pages} {034101} (\bibinfo {year}
  {2015})}\BibitemShut {NoStop}%
\bibitem [{\citenamefont {Semenova}\ \emph {et~al.}(2016)\citenamefont
  {Semenova}, \citenamefont {Zakharova}, \citenamefont {Anishchenko},\ and\
  \citenamefont {Sch{\"o}ll}}]{semenova2016coherence}%
  \BibitemOpen
  \bibfield  {author} {\bibinfo {author} {\bibfnamefont {N.}~\bibnamefont
  {Semenova}}, \bibinfo {author} {\bibfnamefont {A.}~\bibnamefont {Zakharova}},
  \bibinfo {author} {\bibfnamefont {V.}~\bibnamefont {Anishchenko}},\ and\
  \bibinfo {author} {\bibfnamefont {E.}~\bibnamefont {Sch{\"o}ll}},\ }\bibfield
   {title} {\bibinfo {title} {Coherence-resonance chimeras in a network of
  excitable elements},\ }\href@noop {} {\bibfield  {journal} {\bibinfo
  {journal} {Phys. Rev. Lett.}\ }\textbf {\bibinfo {volume} {117}},\ \bibinfo
  {pages} {014102} (\bibinfo {year} {2016})}\BibitemShut {NoStop}%
\bibitem [{\citenamefont {Nicolaou}\ \emph {et~al.}(2017)\citenamefont
  {Nicolaou}, \citenamefont {Riecke},\ and\ \citenamefont
  {Motter}}]{nicolaou2017chimera}%
  \BibitemOpen
  \bibfield  {author} {\bibinfo {author} {\bibfnamefont {Z.~G.}\ \bibnamefont
  {Nicolaou}}, \bibinfo {author} {\bibfnamefont {H.}~\bibnamefont {Riecke}},\
  and\ \bibinfo {author} {\bibfnamefont {A.~E.}\ \bibnamefont {Motter}},\
  }\bibfield  {title} {\bibinfo {title} {Chimera states in continuous media:
  Existence and distinctness},\ }\href@noop {} {\bibfield  {journal} {\bibinfo
  {journal} {Phys. Rev. Lett.}\ }\textbf {\bibinfo {volume} {119}},\ \bibinfo
  {pages} {244101} (\bibinfo {year} {2017})}\BibitemShut {NoStop}%
\bibitem [{\citenamefont {Kotwal}\ \emph {et~al.}(2017)\citenamefont {Kotwal},
  \citenamefont {Jiang},\ and\ \citenamefont {Abrams}}]{kotwal2017connecting}%
  \BibitemOpen
  \bibfield  {author} {\bibinfo {author} {\bibfnamefont {T.}~\bibnamefont
  {Kotwal}}, \bibinfo {author} {\bibfnamefont {X.}~\bibnamefont {Jiang}},\ and\
  \bibinfo {author} {\bibfnamefont {D.~M.}\ \bibnamefont {Abrams}},\ }\bibfield
   {title} {\bibinfo {title} {Connecting the {K}uramoto model and the chimera
  state},\ }\href@noop {} {\bibfield  {journal} {\bibinfo  {journal} {Phys.
  Rev. Lett.}\ }\textbf {\bibinfo {volume} {119}},\ \bibinfo {pages} {264101}
  (\bibinfo {year} {2017})}\BibitemShut {NoStop}%
\bibitem [{\citenamefont {Panaggio}\ \emph {et~al.}(2016)\citenamefont
  {Panaggio}, \citenamefont {Abrams}, \citenamefont {Ashwin},\ and\
  \citenamefont {Laing}}]{panaggio2016chimera}%
  \BibitemOpen
  \bibfield  {author} {\bibinfo {author} {\bibfnamefont {M.~J.}\ \bibnamefont
  {Panaggio}}, \bibinfo {author} {\bibfnamefont {D.~M.}\ \bibnamefont
  {Abrams}}, \bibinfo {author} {\bibfnamefont {P.}~\bibnamefont {Ashwin}},\
  and\ \bibinfo {author} {\bibfnamefont {C.~R.}\ \bibnamefont {Laing}},\
  }\bibfield  {title} {\bibinfo {title} {Chimera states in networks of phase
  oscillators: The case of two small populations},\ }\href@noop {} {\bibfield
  {journal} {\bibinfo  {journal} {Phys. Rev. E}\ }\textbf {\bibinfo {volume}
  {93}},\ \bibinfo {pages} {012218} (\bibinfo {year} {2016})}\BibitemShut
  {NoStop}%
\bibitem [{\citenamefont {Cho}\ \emph {et~al.}(2017)\citenamefont {Cho},
  \citenamefont {Nishikawa},\ and\ \citenamefont {Motter}}]{cho2017stable}%
  \BibitemOpen
  \bibfield  {author} {\bibinfo {author} {\bibfnamefont {Y.~S.}\ \bibnamefont
  {Cho}}, \bibinfo {author} {\bibfnamefont {T.}~\bibnamefont {Nishikawa}},\
  and\ \bibinfo {author} {\bibfnamefont {A.~E.}\ \bibnamefont {Motter}},\
  }\bibfield  {title} {\bibinfo {title} {Stable chimeras and independently
  synchronizable clusters},\ }\href@noop {} {\bibfield  {journal} {\bibinfo
  {journal} {Phys. Rev. Lett.}\ }\textbf {\bibinfo {volume} {119}},\ \bibinfo
  {pages} {084101} (\bibinfo {year} {2017})}\BibitemShut {NoStop}%
\bibitem [{\citenamefont {Sieber}\ \emph {et~al.}(2014)\citenamefont {Sieber},
  \citenamefont {Omel'chenko},\ and\ \citenamefont
  {Wolfrum}}]{sieber2014controlling}%
  \BibitemOpen
  \bibfield  {author} {\bibinfo {author} {\bibfnamefont {J.}~\bibnamefont
  {Sieber}}, \bibinfo {author} {\bibfnamefont {O.~E.}\ \bibnamefont
  {Omel'chenko}},\ and\ \bibinfo {author} {\bibfnamefont {M.}~\bibnamefont
  {Wolfrum}},\ }\bibfield  {title} {\bibinfo {title} {Controlling unstable
  chaos: stabilizing chimera states by feedback},\ }\href@noop {} {\bibfield
  {journal} {\bibinfo  {journal} {Phys. Rev. Lett.}\ }\textbf {\bibinfo
  {volume} {112}},\ \bibinfo {pages} {054102} (\bibinfo {year}
  {2014})}\BibitemShut {NoStop}%
\bibitem [{\citenamefont {Zhou}\ \emph {et~al.}(2002)\citenamefont {Zhou},
  \citenamefont {Kurths}, \citenamefont {Kiss},\ and\ \citenamefont
  {Hudson}}]{zhou2002noise}%
  \BibitemOpen
  \bibfield  {author} {\bibinfo {author} {\bibfnamefont {C.}~\bibnamefont
  {Zhou}}, \bibinfo {author} {\bibfnamefont {J.}~\bibnamefont {Kurths}},
  \bibinfo {author} {\bibfnamefont {I.~Z.}\ \bibnamefont {Kiss}},\ and\
  \bibinfo {author} {\bibfnamefont {J.~L.}\ \bibnamefont {Hudson}},\ }\bibfield
   {title} {\bibinfo {title} {Noise-enhanced phase synchronization of chaotic
  oscillators},\ }\href@noop {} {\bibfield  {journal} {\bibinfo  {journal}
  {Phys. Rev. Lett.}\ }\textbf {\bibinfo {volume} {89}},\ \bibinfo {pages}
  {014101} (\bibinfo {year} {2002})}\BibitemShut {NoStop}%
\bibitem [{\citenamefont {Goldobin}\ and\ \citenamefont
  {Pikovsky}(2005)}]{goldobin2005synchronization}%
  \BibitemOpen
  \bibfield  {author} {\bibinfo {author} {\bibfnamefont {D.~S.}\ \bibnamefont
  {Goldobin}}\ and\ \bibinfo {author} {\bibfnamefont {A.~S.}\ \bibnamefont
  {Pikovsky}},\ }\bibfield  {title} {\bibinfo {title} {Synchronization of
  self-sustained oscillators by common white noise},\ }\href@noop {} {\bibfield
   {journal} {\bibinfo  {journal} {Physica A}\ }\textbf {\bibinfo {volume}
  {351}},\ \bibinfo {pages} {126} (\bibinfo {year} {2005})}\BibitemShut
  {NoStop}%
\bibitem [{\citenamefont {Nagai}\ and\ \citenamefont
  {Kori}(2010)}]{nagai2010noise}%
  \BibitemOpen
  \bibfield  {author} {\bibinfo {author} {\bibfnamefont {K.~H.}\ \bibnamefont
  {Nagai}}\ and\ \bibinfo {author} {\bibfnamefont {H.}~\bibnamefont {Kori}},\
  }\bibfield  {title} {\bibinfo {title} {Noise-induced synchronization of a
  large population of globally coupled nonidentical oscillators},\ }\href@noop
  {} {\bibfield  {journal} {\bibinfo  {journal} {Phys. Rev. E}\ }\textbf
  {\bibinfo {volume} {81}},\ \bibinfo {pages} {065202} (\bibinfo {year}
  {2010})}\BibitemShut {NoStop}%
\bibitem [{\citenamefont {Pimenova}\ \emph {et~al.}(2016)\citenamefont
  {Pimenova}, \citenamefont {Goldobin}, \citenamefont {Rosenblum},\ and\
  \citenamefont {Pikovsky}}]{pimenova2016interplay}%
  \BibitemOpen
  \bibfield  {author} {\bibinfo {author} {\bibfnamefont {A.~V.}\ \bibnamefont
  {Pimenova}}, \bibinfo {author} {\bibfnamefont {D.~S.}\ \bibnamefont
  {Goldobin}}, \bibinfo {author} {\bibfnamefont {M.}~\bibnamefont
  {Rosenblum}},\ and\ \bibinfo {author} {\bibfnamefont {A.}~\bibnamefont
  {Pikovsky}},\ }\bibfield  {title} {\bibinfo {title} {Interplay of coupling
  and common noise at the transition to synchrony in oscillator populations},\
  }\href@noop {} {\bibfield  {journal} {\bibinfo  {journal} {Sci. Rep.}\
  }\textbf {\bibinfo {volume} {6}},\ \bibinfo {pages} {38518} (\bibinfo {year}
  {2016})}\BibitemShut {NoStop}%
\bibitem [{git()}]{github}%
  \BibitemOpen
  \href@noop {} {}\bibinfo {note} {Our Julia code to interactively explore
  chimera dynamics is available at
  \url{https://github.com/y-z-zhang/chimera_mechanism}.}\BibitemShut {Stop}%
\bibitem [{\citenamefont {Belykh}\ and\ \citenamefont
  {Hasler}(2011)}]{belykh2011mesoscale}%
  \BibitemOpen
  \bibfield  {author} {\bibinfo {author} {\bibfnamefont {I.}~\bibnamefont
  {Belykh}}\ and\ \bibinfo {author} {\bibfnamefont {M.}~\bibnamefont
  {Hasler}},\ }\bibfield  {title} {\bibinfo {title} {Mesoscale and clusters of
  synchrony in networks of bursting neurons},\ }\href@noop {} {\bibfield
  {journal} {\bibinfo  {journal} {Chaos}\ }\textbf {\bibinfo {volume} {21}},\
  \bibinfo {pages} {016106} (\bibinfo {year} {2011})}\BibitemShut {NoStop}%
\bibitem [{\citenamefont {Pecora}\ \emph {et~al.}(2014)\citenamefont {Pecora},
  \citenamefont {Sorrentino}, \citenamefont {Hagerstrom}, \citenamefont
  {Murphy},\ and\ \citenamefont {Roy}}]{pecora2014cluster}%
  \BibitemOpen
  \bibfield  {author} {\bibinfo {author} {\bibfnamefont {L.~M.}\ \bibnamefont
  {Pecora}}, \bibinfo {author} {\bibfnamefont {F.}~\bibnamefont {Sorrentino}},
  \bibinfo {author} {\bibfnamefont {A.~M.}\ \bibnamefont {Hagerstrom}},
  \bibinfo {author} {\bibfnamefont {T.~E.}\ \bibnamefont {Murphy}},\ and\
  \bibinfo {author} {\bibfnamefont {R.}~\bibnamefont {Roy}},\ }\bibfield
  {title} {\bibinfo {title} {Cluster synchronization and isolated
  desynchronization in complex networks with symmetries},\ }\href@noop {}
  {\bibfield  {journal} {\bibinfo  {journal} {Nat. Commun.}\ }\textbf {\bibinfo
  {volume} {5}},\ \bibinfo {pages} {4079} (\bibinfo {year} {2014})}\BibitemShut
  {NoStop}%
\bibitem [{\citenamefont {Schaub}\ \emph {et~al.}(2016)\citenamefont {Schaub},
  \citenamefont {O'Clery}, \citenamefont {Billeh}, \citenamefont {Delvenne},
  \citenamefont {Lambiotte},\ and\ \citenamefont {Barahona}}]{schaub2016graph}%
  \BibitemOpen
  \bibfield  {author} {\bibinfo {author} {\bibfnamefont {M.~T.}\ \bibnamefont
  {Schaub}}, \bibinfo {author} {\bibfnamefont {N.}~\bibnamefont {O'Clery}},
  \bibinfo {author} {\bibfnamefont {Y.~N.}\ \bibnamefont {Billeh}}, \bibinfo
  {author} {\bibfnamefont {J.-C.}\ \bibnamefont {Delvenne}}, \bibinfo {author}
  {\bibfnamefont {R.}~\bibnamefont {Lambiotte}},\ and\ \bibinfo {author}
  {\bibfnamefont {M.}~\bibnamefont {Barahona}},\ }\bibfield  {title} {\bibinfo
  {title} {Graph partitions and cluster synchronization in networks of
  oscillators},\ }\href@noop {} {\bibfield  {journal} {\bibinfo  {journal}
  {Chaos}\ }\textbf {\bibinfo {volume} {26}},\ \bibinfo {pages} {094821}
  (\bibinfo {year} {2016})}\BibitemShut {NoStop}%
\bibitem [{\citenamefont {Nishikawa}\ and\ \citenamefont
  {Motter}(2006)}]{nishikawa2006maximum}%
  \BibitemOpen
  \bibfield  {author} {\bibinfo {author} {\bibfnamefont {T.}~\bibnamefont
  {Nishikawa}}\ and\ \bibinfo {author} {\bibfnamefont {A.~E.}\ \bibnamefont
  {Motter}},\ }\bibfield  {title} {\bibinfo {title} {Maximum performance at
  minimum cost in network synchronization},\ }\href@noop {} {\bibfield
  {journal} {\bibinfo  {journal} {Physica D}\ }\textbf {\bibinfo {volume}
  {224}},\ \bibinfo {pages} {77} (\bibinfo {year} {2006})}\BibitemShut
  {NoStop}%
\bibitem [{\citenamefont {Hart}\ \emph {et~al.}(2019)\citenamefont {Hart},
  \citenamefont {Zhang}, \citenamefont {Roy},\ and\ \citenamefont
  {Motter}}]{hart2019topological}%
  \BibitemOpen
  \bibfield  {author} {\bibinfo {author} {\bibfnamefont {J.~D.}\ \bibnamefont
  {Hart}}, \bibinfo {author} {\bibfnamefont {Y.}~\bibnamefont {Zhang}},
  \bibinfo {author} {\bibfnamefont {R.}~\bibnamefont {Roy}},\ and\ \bibinfo
  {author} {\bibfnamefont {A.~E.}\ \bibnamefont {Motter}},\ }\bibfield  {title}
  {\bibinfo {title} {Topological control of synchronization patterns: Trading
  symmetry for stability},\ }\href@noop {} {\bibfield  {journal} {\bibinfo
  {journal} {Phys. Rev. Lett.}\ }\textbf {\bibinfo {volume} {122}},\ \bibinfo
  {pages} {058301} (\bibinfo {year} {2019})}\BibitemShut {NoStop}%
\bibitem [{SM1()}]{SM1}%
  \BibitemOpen
  \href@noop {} {}\bibinfo {note} {In Fig.~2(c), $q$ is varied according to
  $\beta = 1 + 4\times 10^{-5}t$, $K = 0.8 + 10^{-4}t$ for $10^{4}$
  iterations.}\BibitemShut {Stop}%
\bibitem [{SM()}]{SM}%
  \BibitemOpen
  \href@noop {} {}\bibinfo {note} {See Supplemental Material at
  \url{http://link.aps.org/supplemental/10.1103/PhysRevLett.126.094101} for
  more examples of strong chimeras, their robustness against oscillator
  heterogeneity, characterization of the desynchronization transition, and more
  details on the effective input.}\BibitemShut {Stop}%
\bibitem [{\citenamefont {Hart}\ \emph {et~al.}(2017)\citenamefont {Hart},
  \citenamefont {Schmadel}, \citenamefont {Murphy},\ and\ \citenamefont
  {Roy}}]{hart2017experiments}%
  \BibitemOpen
  \bibfield  {author} {\bibinfo {author} {\bibfnamefont {J.~D.}\ \bibnamefont
  {Hart}}, \bibinfo {author} {\bibfnamefont {D.~C.}\ \bibnamefont {Schmadel}},
  \bibinfo {author} {\bibfnamefont {T.~E.}\ \bibnamefont {Murphy}},\ and\
  \bibinfo {author} {\bibfnamefont {R.}~\bibnamefont {Roy}},\ }\bibfield
  {title} {\bibinfo {title} {Experiments with arbitrary networks in
  time-multiplexed delay systems},\ }\href@noop {} {\bibfield  {journal}
  {\bibinfo  {journal} {Chaos}\ }\textbf {\bibinfo {volume} {27}},\ \bibinfo
  {pages} {121103} (\bibinfo {year} {2017})}\BibitemShut {NoStop}%
\bibitem [{\citenamefont {Pecora}\ and\ \citenamefont
  {Carroll}(1998)}]{pecora1998master}%
  \BibitemOpen
  \bibfield  {author} {\bibinfo {author} {\bibfnamefont {L.~M.}\ \bibnamefont
  {Pecora}}\ and\ \bibinfo {author} {\bibfnamefont {T.~L.}\ \bibnamefont
  {Carroll}},\ }\bibfield  {title} {\bibinfo {title} {Master stability
  functions for synchronized coupled systems},\ }\href@noop {} {\bibfield
  {journal} {\bibinfo  {journal} {Phys. Rev. Lett.}\ }\textbf {\bibinfo
  {volume} {80}},\ \bibinfo {pages} {2109} (\bibinfo {year}
  {1998})}\BibitemShut {NoStop}%
\bibitem [{\citenamefont {Zakharova}\ \emph {et~al.}(2014)\citenamefont
  {Zakharova}, \citenamefont {Kapeller},\ and\ \citenamefont
  {Sch{\"o}ll}}]{zakharova2014chimera}%
  \BibitemOpen
  \bibfield  {author} {\bibinfo {author} {\bibfnamefont {A.}~\bibnamefont
  {Zakharova}}, \bibinfo {author} {\bibfnamefont {M.}~\bibnamefont
  {Kapeller}},\ and\ \bibinfo {author} {\bibfnamefont {E.}~\bibnamefont
  {Sch{\"o}ll}},\ }\bibfield  {title} {\bibinfo {title} {Chimera death:
  Symmetry breaking in dynamical networks},\ }\href@noop {} {\bibfield
  {journal} {\bibinfo  {journal} {Phys. Rev. Lett.}\ }\textbf {\bibinfo
  {volume} {112}},\ \bibinfo {pages} {154101} (\bibinfo {year}
  {2014})}\BibitemShut {NoStop}%
\bibitem [{\citenamefont {Omelchenko}\ \emph {et~al.}(2011)\citenamefont
  {Omelchenko}, \citenamefont {Maistrenko}, \citenamefont {H{\"o}vel},\ and\
  \citenamefont {Sch{\"o}ll}}]{omelchenko2011loss}%
  \BibitemOpen
  \bibfield  {author} {\bibinfo {author} {\bibfnamefont {I.}~\bibnamefont
  {Omelchenko}}, \bibinfo {author} {\bibfnamefont {Y.}~\bibnamefont
  {Maistrenko}}, \bibinfo {author} {\bibfnamefont {P.}~\bibnamefont
  {H{\"o}vel}},\ and\ \bibinfo {author} {\bibfnamefont {E.}~\bibnamefont
  {Sch{\"o}ll}},\ }\bibfield  {title} {\bibinfo {title} {Loss of coherence in
  dynamical networks: Spatial chaos and chimera states},\ }\href@noop {}
  {\bibfield  {journal} {\bibinfo  {journal} {Phys. Rev. Lett.}\ }\textbf
  {\bibinfo {volume} {106}},\ \bibinfo {pages} {234102} (\bibinfo {year}
  {2011})}\BibitemShut {NoStop}%
\bibitem [{\citenamefont {Omelchenko}\ \emph {et~al.}(2012)\citenamefont
  {Omelchenko}, \citenamefont {Riemenschneider}, \citenamefont {H{\"o}vel},
  \citenamefont {Maistrenko},\ and\ \citenamefont
  {Sch{\"o}ll}}]{omelchenko2012transition}%
  \BibitemOpen
  \bibfield  {author} {\bibinfo {author} {\bibfnamefont {I.}~\bibnamefont
  {Omelchenko}}, \bibinfo {author} {\bibfnamefont {B.}~\bibnamefont
  {Riemenschneider}}, \bibinfo {author} {\bibfnamefont {P.}~\bibnamefont
  {H{\"o}vel}}, \bibinfo {author} {\bibfnamefont {Y.}~\bibnamefont
  {Maistrenko}},\ and\ \bibinfo {author} {\bibfnamefont {E.}~\bibnamefont
  {Sch{\"o}ll}},\ }\bibfield  {title} {\bibinfo {title} {Transition from
  spatial coherence to incoherence in coupled chaotic systems},\ }\href@noop {}
  {\bibfield  {journal} {\bibinfo  {journal} {Phys. Rev. E}\ }\textbf {\bibinfo
  {volume} {85}},\ \bibinfo {pages} {026212} (\bibinfo {year}
  {2012})}\BibitemShut {NoStop}%
\bibitem [{\citenamefont {Jiruska}\ \emph {et~al.}(2013)\citenamefont
  {Jiruska}, \citenamefont {De~Curtis}, \citenamefont {Jefferys}, \citenamefont
  {Schevon}, \citenamefont {Schiff},\ and\ \citenamefont
  {Schindler}}]{jiruska2013synchronization}%
  \BibitemOpen
  \bibfield  {author} {\bibinfo {author} {\bibfnamefont {P.}~\bibnamefont
  {Jiruska}}, \bibinfo {author} {\bibfnamefont {M.}~\bibnamefont {De~Curtis}},
  \bibinfo {author} {\bibfnamefont {J.~G.}\ \bibnamefont {Jefferys}}, \bibinfo
  {author} {\bibfnamefont {C.~A.}\ \bibnamefont {Schevon}}, \bibinfo {author}
  {\bibfnamefont {S.~J.}\ \bibnamefont {Schiff}},\ and\ \bibinfo {author}
  {\bibfnamefont {K.}~\bibnamefont {Schindler}},\ }\bibfield  {title} {\bibinfo
  {title} {Synchronization and desynchronization in epilepsy: Controversies and
  hypotheses},\ }\href@noop {} {\bibfield  {journal} {\bibinfo  {journal} {J.
  Physiol.}\ }\textbf {\bibinfo {volume} {591}},\ \bibinfo {pages} {787}
  (\bibinfo {year} {2013})}\BibitemShut {NoStop}%
\bibitem [{\citenamefont {Rothkegel}\ and\ \citenamefont
  {Lehnertz}(2014)}]{rothkegel2014irregular}%
  \BibitemOpen
  \bibfield  {author} {\bibinfo {author} {\bibfnamefont {A.}~\bibnamefont
  {Rothkegel}}\ and\ \bibinfo {author} {\bibfnamefont {K.}~\bibnamefont
  {Lehnertz}},\ }\bibfield  {title} {\bibinfo {title} {Irregular macroscopic
  dynamics due to chimera states in small-world networks of pulse-coupled
  oscillators},\ }\href@noop {} {\bibfield  {journal} {\bibinfo  {journal} {New
  J. Phys.}\ }\textbf {\bibinfo {volume} {16}},\ \bibinfo {pages} {055006}
  (\bibinfo {year} {2014})}\BibitemShut {NoStop}%
\bibitem [{\citenamefont {Andrzejak}\ \emph {et~al.}(2016)\citenamefont
  {Andrzejak}, \citenamefont {Rummel}, \citenamefont {Mormann},\ and\
  \citenamefont {Schindler}}]{andrzejak2016all}%
  \BibitemOpen
  \bibfield  {author} {\bibinfo {author} {\bibfnamefont {R.~G.}\ \bibnamefont
  {Andrzejak}}, \bibinfo {author} {\bibfnamefont {C.}~\bibnamefont {Rummel}},
  \bibinfo {author} {\bibfnamefont {F.}~\bibnamefont {Mormann}},\ and\ \bibinfo
  {author} {\bibfnamefont {K.}~\bibnamefont {Schindler}},\ }\bibfield  {title}
  {\bibinfo {title} {All together now: Analogies between chimera state
  collapses and epileptic seizures},\ }\href@noop {} {\bibfield  {journal}
  {\bibinfo  {journal} {Sci. Rep.}\ }\textbf {\bibinfo {volume} {6}},\ \bibinfo
  {pages} {23000} (\bibinfo {year} {2016})}\BibitemShut {NoStop}%
\bibitem [{\citenamefont {Chouzouris}\ \emph {et~al.}(2018)\citenamefont
  {Chouzouris}, \citenamefont {Omelchenko}, \citenamefont {Zakharova},
  \citenamefont {Hlinka}, \citenamefont {Jiruska},\ and\ \citenamefont
  {Sch{\"o}ll}}]{chouzouris2018chimera}%
  \BibitemOpen
  \bibfield  {author} {\bibinfo {author} {\bibfnamefont {T.}~\bibnamefont
  {Chouzouris}}, \bibinfo {author} {\bibfnamefont {I.}~\bibnamefont
  {Omelchenko}}, \bibinfo {author} {\bibfnamefont {A.}~\bibnamefont
  {Zakharova}}, \bibinfo {author} {\bibfnamefont {J.}~\bibnamefont {Hlinka}},
  \bibinfo {author} {\bibfnamefont {P.}~\bibnamefont {Jiruska}},\ and\ \bibinfo
  {author} {\bibfnamefont {E.}~\bibnamefont {Sch{\"o}ll}},\ }\bibfield  {title}
  {\bibinfo {title} {Chimera states in brain networks: Empirical neural vs.
  modular fractal connectivity},\ }\href@noop {} {\bibfield  {journal}
  {\bibinfo  {journal} {Chaos}\ }\textbf {\bibinfo {volume} {28}},\ \bibinfo
  {pages} {045112} (\bibinfo {year} {2018})}\BibitemShut {NoStop}%
\bibitem [{\citenamefont {Lainscsek}\ \emph {et~al.}(2019)\citenamefont
  {Lainscsek}, \citenamefont {Rungratsameetaweemana}, \citenamefont {Cash},\
  and\ \citenamefont {Sejnowski}}]{lainscsek2019cortical}%
  \BibitemOpen
  \bibfield  {author} {\bibinfo {author} {\bibfnamefont {C.}~\bibnamefont
  {Lainscsek}}, \bibinfo {author} {\bibfnamefont {N.}~\bibnamefont
  {Rungratsameetaweemana}}, \bibinfo {author} {\bibfnamefont {S.~S.}\
  \bibnamefont {Cash}},\ and\ \bibinfo {author} {\bibfnamefont {T.~J.}\
  \bibnamefont {Sejnowski}},\ }\bibfield  {title} {\bibinfo {title} {Cortical
  chimera states predict epileptic seizures},\ }\href@noop {} {\bibfield
  {journal} {\bibinfo  {journal} {Chaos}\ }\textbf {\bibinfo {volume} {29}},\
  \bibinfo {pages} {121106} (\bibinfo {year} {2019})}\BibitemShut {NoStop}%
\bibitem [{\citenamefont {Asadi-Pooya}\ \emph {et~al.}(2008)\citenamefont
  {Asadi-Pooya}, \citenamefont {Sharan}, \citenamefont {Nei},\ and\
  \citenamefont {Sperling}}]{asadi2008corpus}%
  \BibitemOpen
  \bibfield  {author} {\bibinfo {author} {\bibfnamefont {A.~A.}\ \bibnamefont
  {Asadi-Pooya}}, \bibinfo {author} {\bibfnamefont {A.}~\bibnamefont {Sharan}},
  \bibinfo {author} {\bibfnamefont {M.}~\bibnamefont {Nei}},\ and\ \bibinfo
  {author} {\bibfnamefont {M.~R.}\ \bibnamefont {Sperling}},\ }\bibfield
  {title} {\bibinfo {title} {Corpus callosotomy},\ }\href@noop {} {\bibfield
  {journal} {\bibinfo  {journal} {Epilepsy Behav.}\ }\textbf {\bibinfo {volume}
  {13}},\ \bibinfo {pages} {271} (\bibinfo {year} {2008})}\BibitemShut
  {NoStop}%
\bibitem [{\citenamefont {Netoff}\ and\ \citenamefont
  {Schiff}(2002)}]{netoff2002decreased}%
  \BibitemOpen
  \bibfield  {author} {\bibinfo {author} {\bibfnamefont {T.~I.}\ \bibnamefont
  {Netoff}}\ and\ \bibinfo {author} {\bibfnamefont {S.~J.}\ \bibnamefont
  {Schiff}},\ }\bibfield  {title} {\bibinfo {title} {Decreased neuronal
  synchronization during experimental seizures},\ }\href@noop {} {\bibfield
  {journal} {\bibinfo  {journal} {J. Neurosci.}\ }\textbf {\bibinfo {volume}
  {22}},\ \bibinfo {pages} {7297} (\bibinfo {year} {2002})}\BibitemShut
  {NoStop}%
\bibitem [{\citenamefont {Mormann}\ \emph {et~al.}(2003)\citenamefont
  {Mormann}, \citenamefont {Kreuz}, \citenamefont {Andrzejak}, \citenamefont
  {David}, \citenamefont {Lehnertz},\ and\ \citenamefont
  {Elger}}]{mormann2003epileptic}%
  \BibitemOpen
  \bibfield  {author} {\bibinfo {author} {\bibfnamefont {F.}~\bibnamefont
  {Mormann}}, \bibinfo {author} {\bibfnamefont {T.}~\bibnamefont {Kreuz}},
  \bibinfo {author} {\bibfnamefont {R.~G.}\ \bibnamefont {Andrzejak}}, \bibinfo
  {author} {\bibfnamefont {P.}~\bibnamefont {David}}, \bibinfo {author}
  {\bibfnamefont {K.}~\bibnamefont {Lehnertz}},\ and\ \bibinfo {author}
  {\bibfnamefont {C.~E.}\ \bibnamefont {Elger}},\ }\bibfield  {title} {\bibinfo
  {title} {Epileptic seizures are preceded by a decrease in synchronization},\
  }\href@noop {} {\bibfield  {journal} {\bibinfo  {journal} {Epilepsy Res.}\
  }\textbf {\bibinfo {volume} {53}},\ \bibinfo {pages} {173} (\bibinfo {year}
  {2003})}\BibitemShut {NoStop}%
\bibitem [{\citenamefont {Schindler}\ \emph {et~al.}(2007)\citenamefont
  {Schindler}, \citenamefont {Leung}, \citenamefont {Elger},\ and\
  \citenamefont {Lehnertz}}]{schindler2007assessing}%
  \BibitemOpen
  \bibfield  {author} {\bibinfo {author} {\bibfnamefont {K.}~\bibnamefont
  {Schindler}}, \bibinfo {author} {\bibfnamefont {H.}~\bibnamefont {Leung}},
  \bibinfo {author} {\bibfnamefont {C.~E.}\ \bibnamefont {Elger}},\ and\
  \bibinfo {author} {\bibfnamefont {K.}~\bibnamefont {Lehnertz}},\ }\bibfield
  {title} {\bibinfo {title} {Assessing seizure dynamics by analysing the
  correlation structure of multichannel intracranial {EEG}},\ }\href@noop {}
  {\bibfield  {journal} {\bibinfo  {journal} {Brain}\ }\textbf {\bibinfo
  {volume} {130}},\ \bibinfo {pages} {65} (\bibinfo {year} {2007})}\BibitemShut
  {NoStop}%
\bibitem [{\citenamefont {Bogomolov}\ \emph {et~al.}(2017)\citenamefont
  {Bogomolov}, \citenamefont {Slepnev}, \citenamefont {Strelkova},
  \citenamefont {Sch{\"o}ll},\ and\ \citenamefont
  {Anishchenko}}]{bogomolov2017mechanisms}%
  \BibitemOpen
  \bibfield  {author} {\bibinfo {author} {\bibfnamefont {S.~A.}\ \bibnamefont
  {Bogomolov}}, \bibinfo {author} {\bibfnamefont {A.~V.}\ \bibnamefont
  {Slepnev}}, \bibinfo {author} {\bibfnamefont {G.~I.}\ \bibnamefont
  {Strelkova}}, \bibinfo {author} {\bibfnamefont {E.}~\bibnamefont
  {Sch{\"o}ll}},\ and\ \bibinfo {author} {\bibfnamefont {V.~S.}\ \bibnamefont
  {Anishchenko}},\ }\bibfield  {title} {\bibinfo {title} {Mechanisms of
  appearance of amplitude and phase chimera states in ensembles of nonlocally
  coupled chaotic systems},\ }\href@noop {} {\bibfield  {journal} {\bibinfo
  {journal} {Commun. Nonlinear Sci. Numer. Simulat.}\ }\textbf {\bibinfo
  {volume} {43}},\ \bibinfo {pages} {25} (\bibinfo {year} {2017})}\BibitemShut
  {NoStop}%
\bibitem [{\citenamefont {Zhou}\ and\ \citenamefont
  {Kurths}(2002)}]{zhou2002noise2}%
  \BibitemOpen
  \bibfield  {author} {\bibinfo {author} {\bibfnamefont {C.}~\bibnamefont
  {Zhou}}\ and\ \bibinfo {author} {\bibfnamefont {J.}~\bibnamefont {Kurths}},\
  }\bibfield  {title} {\bibinfo {title} {Noise-induced phase synchronization
  and synchronization transitions in chaotic oscillators},\ }\href@noop {}
  {\bibfield  {journal} {\bibinfo  {journal} {Phys. Rev. Lett.}\ }\textbf
  {\bibinfo {volume} {88}},\ \bibinfo {pages} {230602} (\bibinfo {year}
  {2002})}\BibitemShut {NoStop}%
\bibitem [{\citenamefont {Teramae}\ and\ \citenamefont
  {Tanaka}(2004)}]{teramae2004robustness}%
  \BibitemOpen
  \bibfield  {author} {\bibinfo {author} {\bibfnamefont {J.-N.}\ \bibnamefont
  {Teramae}}\ and\ \bibinfo {author} {\bibfnamefont {D.}~\bibnamefont
  {Tanaka}},\ }\bibfield  {title} {\bibinfo {title} {Robustness of the
  noise-induced phase synchronization in a general class of limit cycle
  oscillators},\ }\href@noop {} {\bibfield  {journal} {\bibinfo  {journal}
  {Phys. Rev. Lett.}\ }\textbf {\bibinfo {volume} {93}},\ \bibinfo {pages}
  {204103} (\bibinfo {year} {2004})}\BibitemShut {NoStop}%
\bibitem [{\citenamefont {Nakao}\ \emph {et~al.}(2007)\citenamefont {Nakao},
  \citenamefont {Arai},\ and\ \citenamefont {Kawamura}}]{nakao2007noise}%
  \BibitemOpen
  \bibfield  {author} {\bibinfo {author} {\bibfnamefont {H.}~\bibnamefont
  {Nakao}}, \bibinfo {author} {\bibfnamefont {K.}~\bibnamefont {Arai}},\ and\
  \bibinfo {author} {\bibfnamefont {Y.}~\bibnamefont {Kawamura}},\ }\bibfield
  {title} {\bibinfo {title} {Noise-induced synchronization and clustering in
  ensembles of uncoupled limit-cycle oscillators},\ }\href@noop {} {\bibfield
  {journal} {\bibinfo  {journal} {Phys. Rev. Lett.}\ }\textbf {\bibinfo
  {volume} {98}},\ \bibinfo {pages} {184101} (\bibinfo {year}
  {2007})}\BibitemShut {NoStop}%
\bibitem [{\citenamefont {Ullner}\ \emph {et~al.}(2009)\citenamefont {Ullner},
  \citenamefont {Buceta}, \citenamefont {D{\'\i}ez-Noguera},\ and\
  \citenamefont {Garc{\'\i}a-Ojalvo}}]{ullner2009noise}%
  \BibitemOpen
  \bibfield  {author} {\bibinfo {author} {\bibfnamefont {E.}~\bibnamefont
  {Ullner}}, \bibinfo {author} {\bibfnamefont {J.}~\bibnamefont {Buceta}},
  \bibinfo {author} {\bibfnamefont {A.}~\bibnamefont {D{\'\i}ez-Noguera}},\
  and\ \bibinfo {author} {\bibfnamefont {J.}~\bibnamefont
  {Garc{\'\i}a-Ojalvo}},\ }\bibfield  {title} {\bibinfo {title} {Noise-induced
  coherence in multicellular circadian clocks},\ }\href@noop {} {\bibfield
  {journal} {\bibinfo  {journal} {Biophys. J.}\ }\textbf {\bibinfo {volume}
  {96}},\ \bibinfo {pages} {3573} (\bibinfo {year} {2009})}\BibitemShut
  {NoStop}%
\end{thebibliography}%

\end{document}